\documentclass[%
preprint,
amsmath,
amssymb,
aps,
floatfix,
]{revtex4-2}
\usepackage[utf8]{inputenc}
\usepackage[T1]{fontenc}
\usepackage{color}
\usepackage{graphicx}
\usepackage{dcolumn}
\usepackage{bm}
\usepackage{xfrac}
\usepackage{upgreek}

\usepackage{dcolumn}
\usepackage{bm}
\usepackage{braket}

\graphicspath{Figures}

\begin{document}


\title{Coherent light scattering from a telecom C-band quantum dot}
\author{L. Wells\(^{1,2}\)}
\author{T. M\"{u}ller\(^{1}\)}
\email{tina.muller@crl.toshiba.co.uk}
\author{R.M. Stevenson\(^{1}\)}
\author{J. Skiba-Szymanska\(^{1}\)}
\author{D.A. Ritchie\(^{2}\)}
\author{A.J. Shields\(^{1}\)}
\affiliation{\(^{1}\)Toshiba Europe Limited, 208 Science Park, Milton Road, Cambridge, CB4 0GZ, UK}
\affiliation{\(^{2}\)Cavendish Laboratory, University of Cambridge, JJ Thomson Avenue, Cambridge, CB3 0HE, UK}

\date{\today}

\begin{abstract}
\noindent Quantum networks have the potential to transform secure communication via quantum key distribution and enable novel concepts in distributed quantum computing and sensing. Coherent quantum light generation at telecom wavelengths is fundamental for fibre-based network implementations, but Fourier-limited emission and subnatural linewidth photons have so far only been reported from systems operating in the visible to near-infrared wavelength range.  Here, we use InAs/InP quantum dots to demonstrate photons with coherence times much longer than the Fourier limit at telecom wavelength. Evidence of the responsible elastic laser scattering mechanism is observed in a distinct signature in two-photon interference measurements, and is confirmed using a direct measurement of the emission coherence. Further, we show that even the inelastically scattered photons have coherence times within the error bars of the Fourier limit. Finally, we make direct use of the minimal attenuation in fibre for these photons by measuring two-photon interference after 25 km of fibre, thereby demonstrating indistinguishability of photons emitted about 100 000 excitation cycles apart. 
\end{abstract}
\maketitle
\newpage
\section{Introduction}

Establishing long distance quantum networks relies on the efficient exchange of quantum information, conveniently encoded in photonic qubits \cite{Kimble.2008}. Quantum light sources emitting at telecom wavelengths are fundamental to this endeavor, due to the minimal absorption window of standard fibre networks at these wavelengths. As a consequence, there has been a lot of interest recently in developing novel quantum systems with direct emission at these wavelengths, where III-V semiconductor systems ranging from InP to wide bandgap materials such as GaP and SiN \cite{Ward.2014, Dibos.2018, Zhou.2018, Merkel.2020, Wolfowicz.2020, Durand.2021} have shown promise as quantum light sources. For emission in the telecom C-band, quantum dot (QD) technology has been most prominent so far \cite{Vajner.2022}, where two different material systems, modified InAs/GaAs and InAs/InP based, respectively, have been pursued \cite{Portalupi.2019, Anderson.2021}. These systems have made leaps in their development recently, maturing from showing evidence of single photon emission \cite{Kim.2005, Cade.2006} to demonstrations of entangled photon emission \cite{Olbrich.2017, Muller.2018} and the development of a spin-photon interface \cite{Dusanowski.2022}.

A key component for any interference-based quantum network applications is a source of coherent photons. The coherence of the photons is ultimately limited by the radiative linewidth of the underlying transition, where the coherence time $T_2$ cannot exceed twice the radiative lifetime $T_1$ (Fourier limit). However, for solid state emitters, reaching this limit is very challenging due to the inevitable coupling of the emitter to its host matrix and the associated decoherence processes. While resonant excitation is key to minimizing such noise processes \cite{Kuhlmann.2013}, previous demonstrations of Fourier-limited emission from QDs have been limited to lower-wavelength regions around 900 nm and have further relied on cavity enhancement, reducing $T_1$ below the timescale of the dephasing processes \cite{XingDing.2016, Wang.2016, Somaschi.2016}, or on manipulation of the noise processes in the environment \cite{Kuhlmann.2015}. 

To produce photons with coherence times even beyond the Fourier limit, it is possible to take advantage of an elatic scattering process often termed Resonant Raileigh Scattering (RRS), whereby resonant laser light can be elastically scattered from quantum emitters even at excitation powers approaching saturation \cite{Bennett.2016}. First demonstrated in 900-nm QDs in a decade ago \cite{Nguyen.2011, Matthiesen.2012}, this phenomenon continues to be of fundamental interest, and recent work has led to a much improved understanding of the underlying processes \cite{Hanschke.2020, Phillips.2020}. However, the phenomenon has never been observed for telecom wavelength emitters, where arguably it has the highest impact for practical quantum networking applications.

Here, we use the InAs/InP QD platform to demonstrate photons with coherence times much longer than the Fourier limit at telecom wavelength. We first establish resonance fluorescence on a neutral exciton ($X$) transition and characterize the purity of the emission as well as the signal to background ratio achievable in our system. Measuring indistinguishability of consecutively emitted photons, we observe a distinct signature in two-photon correlation traces measurement indicating the presence of photons with coherence times much longer than the Fourier limit. We show that these originate from Resonant Rayleigh scattering rather than residual excitation laser light, and model the observed signatures analytically. The long coherence times are then directly measured in a Michelson interferometer setup. This allows us to show further that even the inelastically scattered part of the photons have coherence times around the Fourier limit. Finally we investigate the indistinguishabilty of photons emitted $\sim$ 100 000 emission cycles apart by sending one of the two photons through a 25-km fibre spool, enabled by the minimal absorption loss experienced by the QD emission near the telecom C-band.

\section{Resonance fluorescence and single photon purity}
\label{sec:g2}

\begin{figure*}
 	\centering
 	\includegraphics[width=1.0\textwidth]{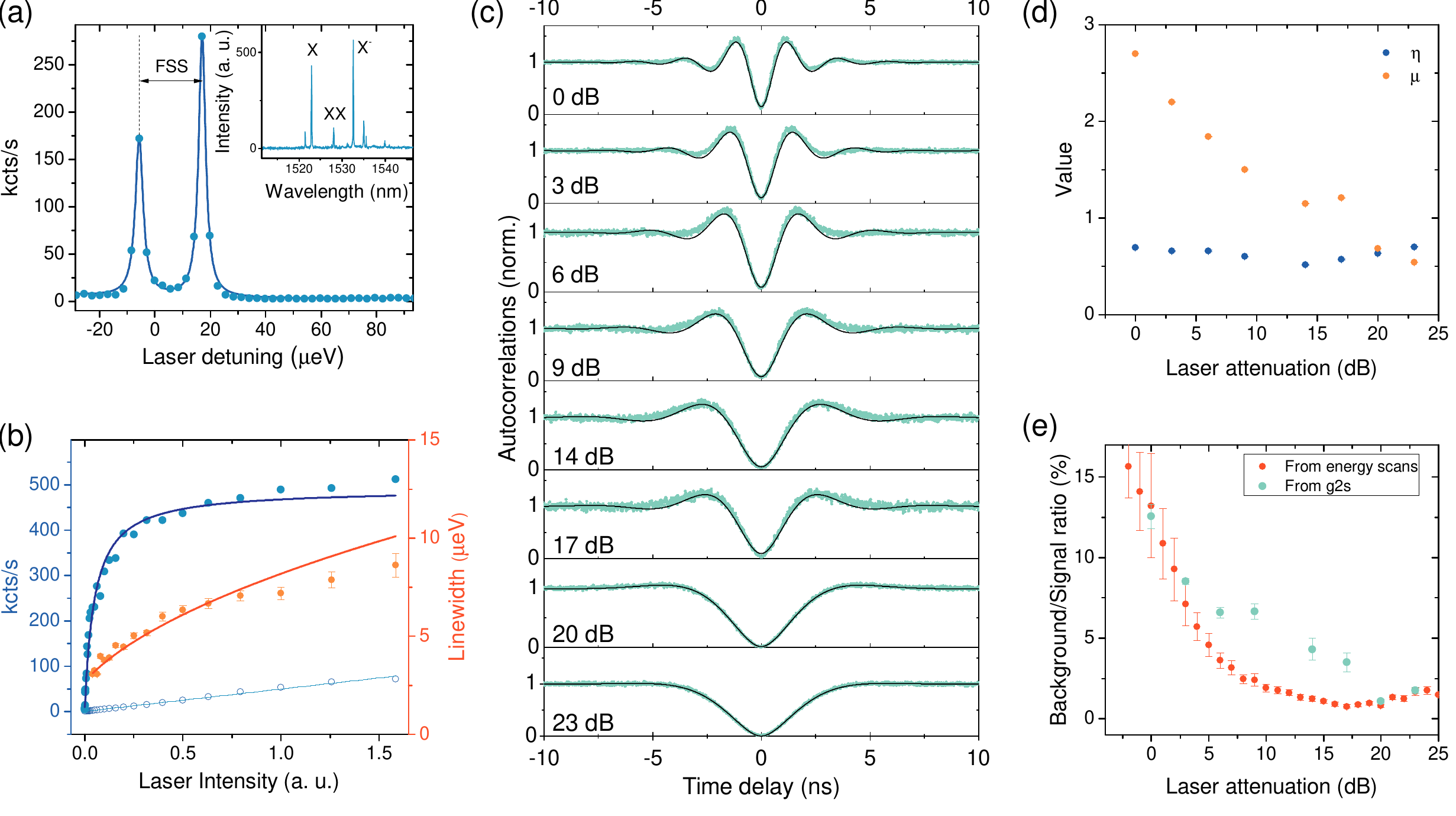}
 	\caption{\textbf{Resonant excitation of a telecom wavelength quantum dot.} (a) Absorption spectrum of a neutrally charged exciton, showing two transitions separated by the fine structure splitting (FSS). The full emission spectrum of this QD, recorded under non-resonant excitation, is shown in the inset. (b) Power-dependent exciation paramters. (c) Autocorrelation measurements.}
 	\label{fig:g2}
 \end{figure*}
 
Our sample consists of InAs/InP quantum dots in a weak planar cavity formed by two Bragg mirrors. For enhanced extraction efficiency, it is further topped by a Zirconia hemisphere. Further details about the sample are given in \ref{app:setup}. Under above-band, non-resonant excitation at 850 nm, we record the typical spectrum from such quantum dots shown in the inset Fig. \ref{fig:g2} (a). To resonantly excite the neutral exciton ($X$) studied here, we tune a narrowband cw laser across the resonance energy and remove any backscattered laser light using polarization suppression, as further described in \ref{app:setup}. Guiding the emission from the quantum dot to superconducting single photon detectors, we observe two transitions separated by the $X$ fine structure splitting of $22.77\pm0.27$ $\upmu$eV, as shown in Fig. \ref{fig:g2} (a). The relative intensities of the transitions is given by their respective overlap with the excitation and detection polarization in our system, which was set to allow both transitions to be visible for maximum efficiency. 

For the remaining measurements described here, we focus on the higher energy transition. Repeating the excitation laser scans as a function of power, we can extract the maximum emission intensity at each power. This results in the count rate saturation behavior clearly seen in Fig. \ref{fig:g2} (b) (blue data points), with a saturation count rate of 491$\pm$9 kcts/s extracted from a fit to the theoretically expected behavior.
Fitting the absorption spectra to a Lorentzian lineshape further allows us to extract the linewidth as a function of power, shown by the orange data points in Fig. \ref{fig:g2} (b). The linewidths can be fitted using the square root dependence on laser intensity expected from pure power broadening, which indicates a natural linewidth of 2.7$\pm$0.12 $\upmu$eV. Further, background emission at the relevant transition energy (consiting of residual laser light as well as detector dark counts and ambient background contributions) can be extrapolated from an empirical polynomial fit to background data only and shows a linear dependence in laser power (Fig. \ref{fig:g2} (b) open circles and cyan curve).

Next, emission is guided to a Hanbury-Brown and Twiss setup for autocorrelation measurements. These are repeated for excitation powers spanning more than two orders of magnitude, as seen in Fig. \ref{fig:g2} (c). All measurements show a pronounced antibunching dip at zero delay. For higher excitation powers, we further observe the onset of Rabi oscillations, which can be fitted using the expected theoretical description
\begin{equation}
g^{(2)}(\tau) = 1-\exp(-\eta |\tau|)\left[\cos{\mu|\tau|} +\frac{\eta}{\mu}\sin{\mu |\tau|}\right],
\label{eq:g2}
\end{equation}
where $\eta = (1/T_1 + 1/T_2)/2$ and $\mu = \sqrt{\Omega^2+\left(\frac{1}{T_1} - \frac{1}{T_2}\right)^2}$. The dependence of these fitting parameters on the excitation power is given in \ref{fig:g2} (d), and shows that $\eta$, which gives the decay of the oscillations, is similar across the different powers, whereas the effective oscillation frequency $\mu$ is reduced with decreasing power as expected. They encompass the three physical quantities $T_1$, the excited state lifetime, $T_2$, the coherence time, and $\Omega$, the Rabi frequency. To determine any two of these, the third one has to be measured independently. In our case, we will measure $T_2$ to extract $T_2$ and $\Omega$ further below.

From the autocorrelation data at zero delay, we can further extract the background/single photon signal ratio under resonant excitation, as shown in Fig. \ref{fig:g2} (e). Values down to 0.01$\pm$0.001 are reached when exciting at 23 dB attenuation. This is about an order of magnitude lower than under non-resonant excitation for these QD \cite{Anderson.2020} and comparable to other work resonantly exciting telecom wavelength QDs \cite{Takemoto.2015, Nawrath.2021}, but does not yet quite reach values reported at lower wavelengths. To investigate where this emission background resulting in non-zero  $g^{(2)}(0)$ values comes from, we compare the $g^{(2)}(0)$ values to the background to signal ratio determined via the Lorentzian fit to the absorption spectrum, which is shown in Fig. \ref{fig:g2} (e) as well. This ratio is dominated by laser breakthrough at high excitation powers, when the QD transition is saturated, and reaches a minimum around 19 dB attenuation. For lower powers, detector dark counts and ambient background become significant. At the high as well as the low end of excitation power, the $g^{(2)}(0)$ values agree very well with the independently determined background ratio, letting us conclude that ambient background and detector dark counts are the biggest contributors to $g^{(2)}(0)$ values at low excitation powers, while at higher powers laser breakthrough constitutes a more significant fraction of the total emission.

\section{Signatures of coherent scattering in two-photon interference measurements}
\begin{figure*}[t]
 	\centering
 	\includegraphics[width=1.0\textwidth]{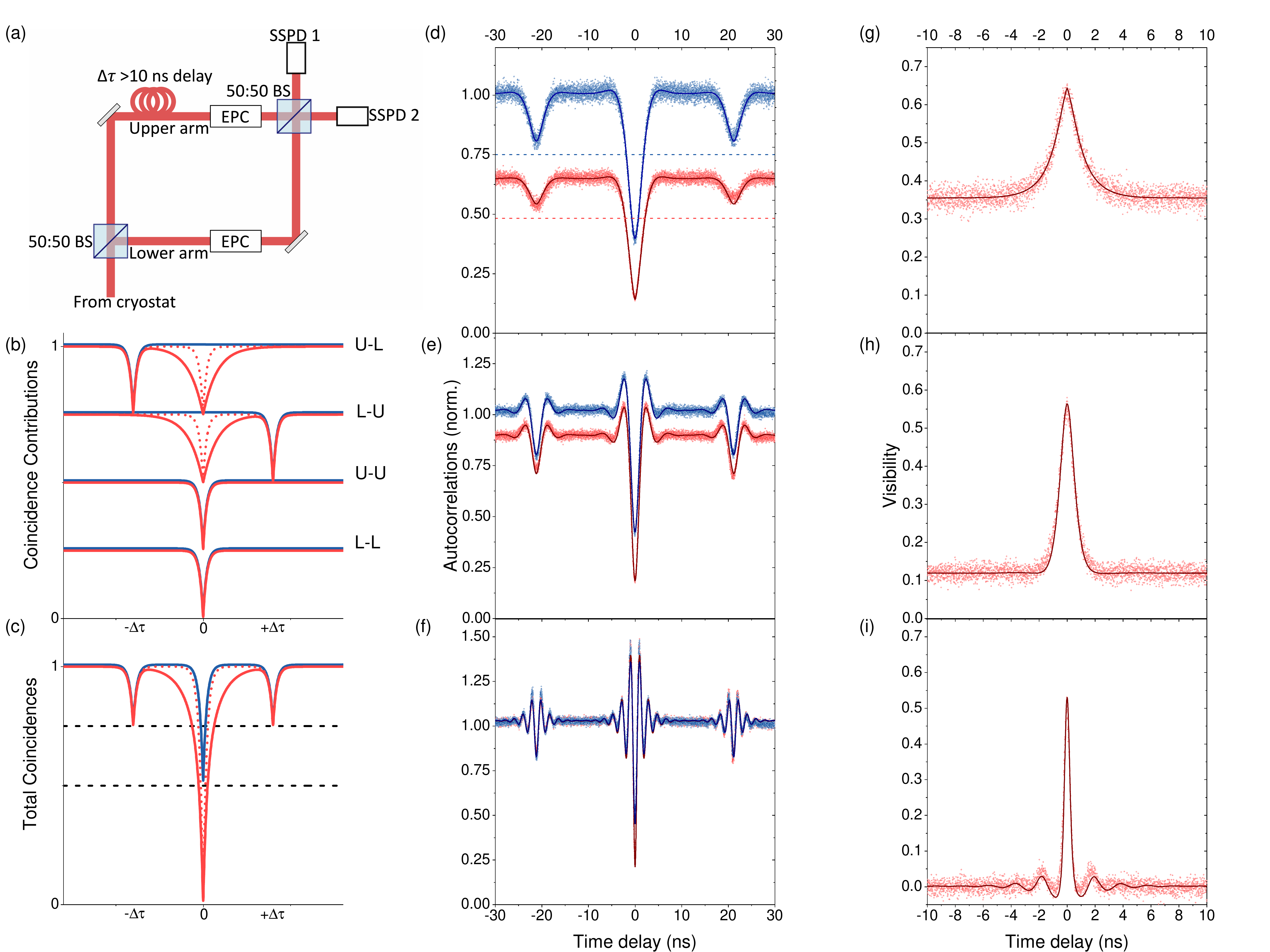}
 	\caption{\textbf{Photon indistinguishability of a resonantly excitation of a telecom wavelength quantum dot.} (a) HOM experimental setup. (b) Schematic showing the individual contributions from the four ways in which coincidences can occur. U (L) indicates the upper (lower) arm was taken for a coincidence event. Blue solid curves stand for cross-polarized coincidences, whereas red solid (dashed) curves stand for co-polarized coincidences where the coherence time was set equal to four times the transition lifetime (the transition lifetime). (c) Schematic showing the total coincidences for cross and co polarized light, with curve colors as above. (d )- (i) Cross (blue) and co (red) polarized normalized autocorrelation measurements for HOM with laser attenuation (d) 20 dB, (e) 9 dB and (f) 0 dB. Corresponding visibility measurements (red) with fits (blue) are also shown for (g) 20 dB, (h) 9 dB and (i) 0 dB.}
 	\label{fig:HOM}
 \end{figure*}

We now investigate the two-photon interference of our QD emission. The collected light is guided to the setup shown in Figure \ref{fig:HOM} (a) for Hong-Ou-Mandel (HOM) type measurements \cite{Hong.1987}. The photons are first separated into two separate arms by a 50:50 beam splitter, and are recombined on a second beamsplitter with an extra delay $\sim$ 20 ns introduced in one of the arms. Performing correlation measurements on the outputs on the superconducting single photon detectors SSPD1 and SSPD2 let us measure the degree of two-photon interference on the second beamsplitter. Electronic polarization controllers (EPCs) are placed in each arm of the setup to control the polarization of the photons, and are set so that the photon polarization is either fully distinguishable (cross-polarized) or fully indistinguishable (co-polarized).

There are four ways in which coincidences can occur, depending on which of the arms the two detected photons traveled though. The individual contributions are illustrated in Figure \ref{fig:HOM} (b), while the combined contributions are shown in Figure \ref{fig:HOM} (c), for the case of a balanced interferometer with equal intensity in both arms. If the two photons detected on SSPD 1 and SSPD 2 both traveled through the same arm (either the upper U or lower L arm in Fig. \ref{fig:HOM} (a) ), an antibunching dip is observed irrespective of the indistinguishability (polarization) of the photons, with the width of the dip determined by the natural lifetime of the transition. This is shown by the lower two curves in \ref{fig:HOM} (b).  If the photons take different paths at the first beamsplitter, the single photon nature of the emission now manifests itself in antibunching dips at $\pm\Delta \tau$. For co-polarized photons, we measure an additional interference dip at zero delay due to the HOM effect, guiding both incoming photons to the same detector and resulting in an absence of coincidences. The width of this additional dip is determined by the coherence time of the arriving photons, which in general differs from the width of the antibunching dip. The overall coincidence pattern, shown in Fig. \ref{fig:HOM} (c), is expected to show a main dip at zero delay, whose depth depends on the indistinguishability of the photons, and two side dips reaching 75\% of the total coincidences for a balanced setup.

Expressing this intuition mathematically, the co-polarized ($\|, \phi=0$) and cross-polarized ($\perp, \phi=\pi/2$) correlations are given by
\begin{equation}\label{eq:corrs}
	g_{\phi}^{(2)}(\tau)=\frac{1}{2}g^{(2)}(\tau)+\frac{1}{2}\left(g^{(2)}(\tau+\Delta\tau)+g^{(2)}(\tau-\Delta\tau)\right)P_{HOM}(\tau, \phi).
\end{equation}
where $\Delta\tau$ is the delay between photons determined by the HOM setup. Here, $P_{HOM}(\tau, \phi)$ describes the two-photon interference probability. For cross-polarized light, $P_{HOM}(\tau, \perp) = 0.5$, and for co-polarized light $P_{HOM}(\tau, \|)$ can be calculated from the known photon mode functions at the second beam splitter in the setup \cite{Legero.2003, Patel.2010, Anderson.2021}, as further detailed in Appendix \ref{app:HOMnorm}.  The resulting visibility is then defined as 
\begin{equation} \label{eq:vis}
	V_{HOM}(\tau)=1-\frac{g_{\parallel}^{(2)}(\tau)}{g_{\perp}^{(2)}(\tau)}.
\end{equation}

The measured correlations for co-and cross-polarized photons are given in Fig. \ref{fig:HOM} (d), (e), and (f) for three different excitation powers. Concentrating on the lowest excitation power in the top panel, the blue data points show a normalized correlation measurement for cross-polarized photons after the HOM setup, where normalization to Poissonian statistics was performed by determining average correlation intensities at large delays. We observe the expected signature with a centre dip just below 50\% and side dips close to the 75\% indicated by the blue dashed line.  Any reduction of this side dip depth, after deconvolution with detector resolution and correction for imbalanced power in the two arms, is due residual excitation laser light. Using the measured dip depth to determine the laser background, we find it to be the order of a few percent for these measurements, in agreement with the direct estimate presented in Fig. \ref{fig:g2} (e). 

The same measurement for co-polarsed photons is shown in red. Here, interestingly, the side dips do not reach 75\% of the local $g^{(2)}(\tau)$ maxima, indicated with the red dashed line. As modeled in detail below, the reason for this reduction in dip depth, which only occurs in the co-polarized data, is two-photon interference of scattered photons with coherence times much longer the $\Delta \tau$. These interference events prevent some of the coincidence events surrounding the side dips, effectively making them appear shallower. The origin of these photons is the resonant Rayleigh scattering process, where a power-dependent fraction of the scattered photons inherit the coherence properties of the excitation laser \cite{Nguyen.2011, Matthiesen.2012, Bennett.2016b}, which has a 10-kHz bandwidth in our case. The presence of these ultralong-coherence-time photons affect the normalization of the co-polarized $g^{(2)}(\tau)$, which is defined as two-photon intensity normalized to the product of the time averaged single photon intensities typically estimated from $g^{(2)}$ at $|\tau| \gg 0$. However, in our measurement, this time average of single photon intensities is underestimated due to interference of the coherently scattered fraction of photons over the entire time delay $\tau$ of our measurement. As discussed below and in Appendix \ref{app:HOMnorm}, by estimating this coherently scattered fraction of the emission from the co-polarized side dip depth, we can compensate for this interference effect to correctly normalize our data and determine $g^{(2)}(0)$ as well as the interference visibility.
 
To confirm the expected power dependence, Fig. \ref{fig:HOM} (d), (e) and (f) show cross- and co-polarized autocorrelation measurements for three different powers. The co-polarized side dips for low powers are decidedly shallower than their cross-polarized counterparts, but increase in depth for increasing power, until they match the cross-polarized side dips at high driving powers. At these powers, we further observe oscillations on either side of the antibunching dips. As in Section \ref{sec:g2}, these are a manifestation of Rabi oscillations. The data are fitted using Eqs. \ref{eq:g2} and \ref{eq:corrs}. We calculate the resulting visibility using Equation \ref{eq:g2} and present the results in Figures \ref{fig:HOM} (g), (h) and (i). Maximum visibility values are 0.64$\pm$0.09, 0.56$\pm$0.05, and 0.53$\pm$0.14 respectively. These values are lower than previously reported results at the same wavelength under resonant excitation \cite{Nawrath.2021} and the result may be surprising given the long coherence times and the good indistinguishability results achieved previously under non-resonant exciation \cite{Anderson.2021}.  It is important to note however that because of the continuous wave excitaiton of our source, these maximum visibility values are not a meaningful measurement of the indistinguishability of the emitted photons \cite{Proux.2015, Baudin.2019, Schofield.2022}, but rather also include the limits of the measurement setup used. The experimental imperfections here are a combination of finite detector resolution, non-ideal branching ratio in the interferometer and imperfect mode overlap in the fibre beamsplitter. For a true measurement of photon indistinguishability, a measurement under pulsed excitation, or an adaptation of Ref. \cite{Schofield.2022} to allow for the coherent fraction of the emission would be needed.

To develop an intuitive understanding of the described signature of photons with ultralong coherence times in correlation measurements, we analytically calculate Equation \ref{eq:corrs}, modeling the incoming field on the second beamsplitter in the HOM setup as the superposition of two traveling waves with coherence times $T_1$ and $T_2$ respectively:
\begin{equation}
	\xi_{1,2}=
	\begin{cases}
	\frac{1}{N}\left(\alpha \sqrt{\frac{2}{T_1}}e^{-\frac{1}{T_1}t-i\omega t}+\beta e^{i\Phi_{1,2}}\sqrt{\frac{2}{T_2}}e^{-\frac{1}{T_2}t-i\omega t}\right), & t\geq0 \\
	0, & t<0
	\end{cases}
	\label{eq:HOMmodel}
\end{equation}
The subscripts refer to the mode function at input 1 and 2 of the second beamsplitter, respectively, and the phases $\Phi_{1,2}$ are random phases included to denote a statistical mixture of the two parts rather than a coherent superposition. The first term refers to spontaneous emission from the QD and has a relatively short coherence time on a nanosecond time scale. This part of the emission has been inelastically scattered during the resonance fluorescence process. The second term denotes photons where the coherence time is inherited from the laser coherence. $|\alpha|^2$ gives the fraction of inelastically scattered photons in the mode, while $|\beta|^2=1-|\alpha|^2$ gives the fraction of elastically scattered photons.
\begin{figure*}[h]
 	\centering
 	\includegraphics[width=0.45\textwidth]{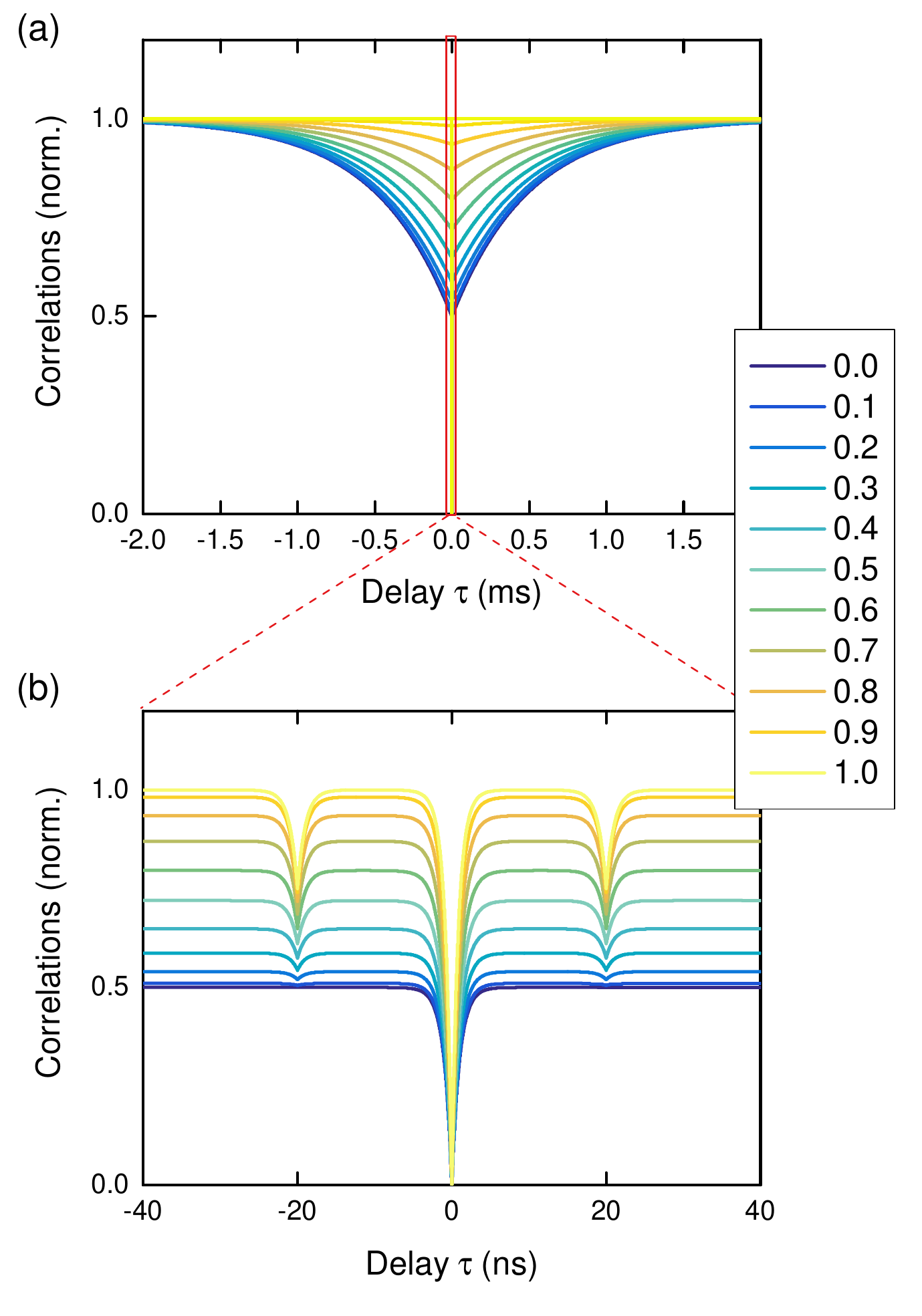}
 	\caption{\textbf{Model of the Hong-Ou-Mandel effect for photons with two coherence times components.} (a) HOM interference on the time scale of the elastically scattered photon coherence time. (b) HOM interference on the time scale of the inelastically scattered photon coherence time.}
 	\label{fig:HOMmodel}
 \end{figure*}

The results from our model are shown in Figure \ref{fig:HOMmodel} (a) for timescales on the order of the laser coherence time. A broad HOM dip is visible due to the elastically scattered fraction of the emitted photons, with the dip depth decreasing with increasing $\alpha$. Looking at the same curves on a ns timescale that is more similar to QD times scales (\ref{fig:HOMmodel} (d)), see that the model reproduces the experimental results shown in Figure \ref{fig:HOM}. If a fraction of the photons have a very long coherence time, it looks as though the dips at$\pm\Delta\tau$ are shallow. Furthermore, the apparent side dip depth is dependent on the fraction of photons emitted via spontaneous emission $|\alpha|^2$. This is the dependency we use to extract the fraction of elastically scattered photons and normalize our co-polarized data (see Appendix \ref{app:HOMnorm}).

\section{Direct measurement of coherence time beyond the Fourier limit}

\begin{figure*}
	\centering
	\includegraphics[width=0.8\textwidth]{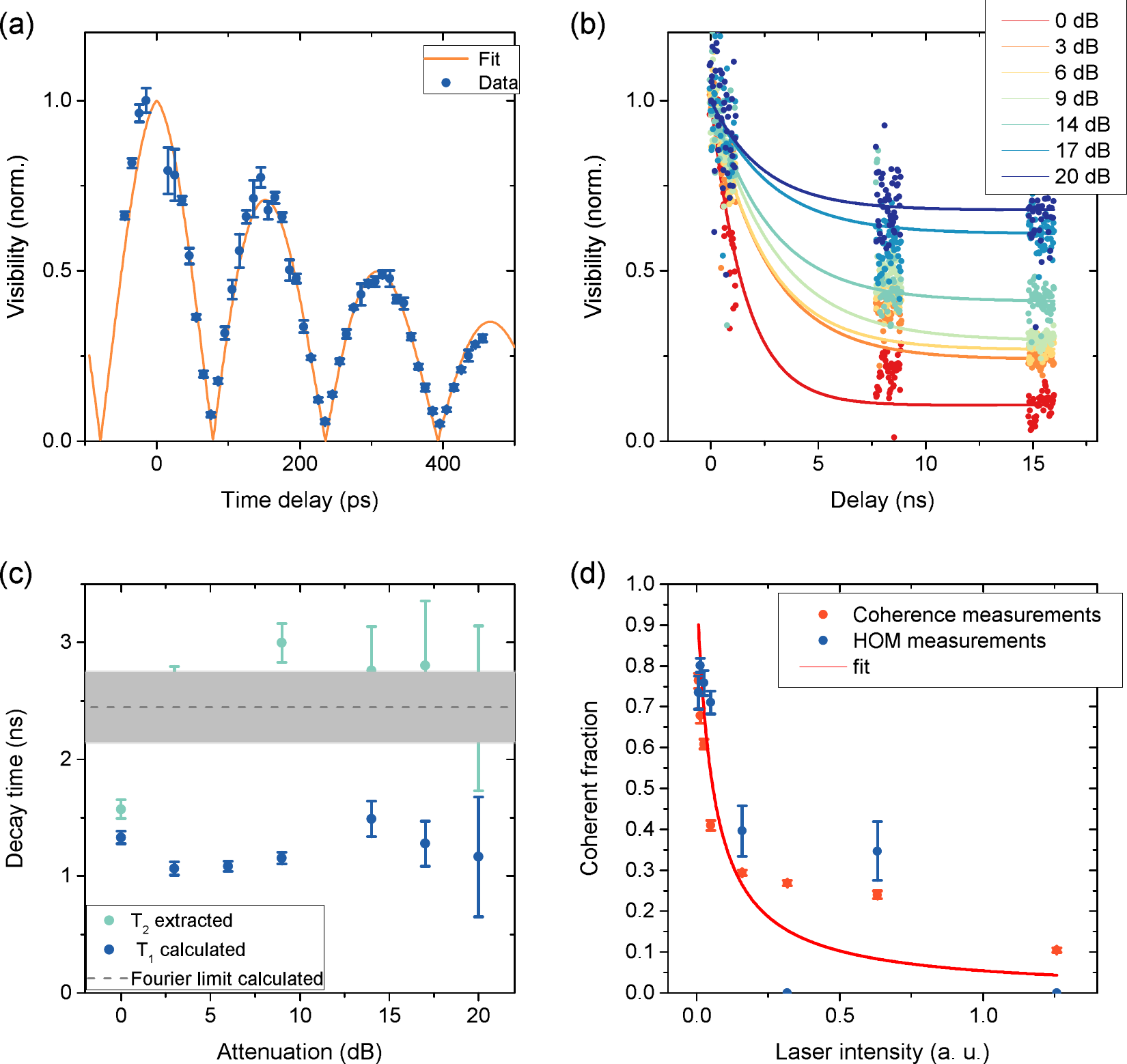}
	\caption{\textbf{Coherence of emission.} (a) Coherence measurement under non-resonant excitation. The visibilities at each delay are determined from sinusoidal fits to interference fringes recorded while changing the relative path lengths over the scale of a few wavelengths using a fibre stretcher, and are normalized to the visibility recorded using a narrowband laser. (b) Coherence measurements as a function of resonant excitation power. Visibilities are determined as in (a). (c) Extracted inelastic coherence times, lifetimes and Fourier limit (see text for details). (d) Coherent fraction of the emission and fit according to text. Note that for the coherent fraction recorded from HOM measurements (blue data points), there are two measurements with identical dip depth for the co- and cross-polarized cases. These measurements give a coherent fraction of zero, and, as becomes clear from the slope of $|\beta|^2(D)$ in Fig. \ref{fig:HOMmaths}, infinite error bars. These error bars were omitted here for clarity.}
	\label{fig:coherence}
\end{figure*}

To obtain direct evidence of the Resonant Rayleigh scattering described above, we measure the coherence time of the photon emission from our QD using a fibre based Michelson interferometer as described in earlier work\cite{Anderson.2021}. To establish a benchmark and compare this QD to our previous results\cite{Anderson.2020, Anderson.2021}, we first measure the emission coherence time under non-resonant cw excitation at 850 nm. The resulting visibility as a function of time delay between the two arms of the interferometer can be seen in Fig. \ref{fig:coherence} (a). It is fitted with the Fourier transform of a double Lorentzian, accounting for the interference between the two fine structure split $X$ states. From the fit, we can extract a coherence time of 447$\pm$15 ps at saturation power, and a fine structure splitting of 26.3$\pm$0.1 $\upmu$eV, which close to the value extracted from the resonant scan in Fig. \ref{fig:g2} (a). 

Next, we record the interference visibility under resonant excitation for resonant laser intensities spanning more than two orders of magnitude. To fully capture the long coherence times expected from the Resonant Rayleigh scattering process, we extend the delay in our Michelson interferometer up to $\sim$ 15 ns. Fig. \ref{fig:coherence} (b) shows the resulting visibilities. We fit the data assuming that the total visibility recorded is the sum of the visibilites from elasitcally and inelastically scattered photons:
\begin{equation}
V_{tot}(\tau) = Ae^{-\frac{t}{\tau_1}}+(1-A)e^{-\frac{t}{\tau_2}},
\label{eq:cohfrac}
\end{equation}
where $A$ represents the coherently scattered fraction of the light, $\tau_1$ is the coherence time of the elastically scattered photons and $\tau_2$ is the coherence time of the inelastically scattered photons corresponding to the natural linewidth of the QD $X$ transition. For the purpose of our fits, $\tau_1$ was set to infinity and the first exponential term therefore set to 1. Looking at Fig. \ref{fig:coherence} (b), we can easily identify the two parts to the visibility in the data: there is an initial exponential decay in coherence on the timescale given by $\tau_2$, followed by a constant section determined by the coherently scattered fraction of the emission. We note that for the highest powers, the data at $\tau \sim$ 7 ns shows markedly higher visibiliy than the data at $\tau \sim$ 15 ns, even though both delays are well beyond the expected Fourier limit and should give similar visibilites originating from only the coherent fraction of the emission. We attribute this to drifts in the experimental setup resulting in a lower effective laser intensity experienced by the quantum dot. To extract the coherent fraction in these cases, we consider the lower possible values by focusing on the data at  $\tau \sim$ 15 ns. Further, as discussed above, there is non-negligible laser breakthrough in our measurements, especially for higher excitation powers. This breakthrough contribution is estimated from the background in the autocorrelation measurements and the data shown in Fig. 2 (b) corrected accordingly before fitting the data with Eq. \ref{eq:cohfrac}.

The QD coherence time values $\tau_2$ extracted from the fits are plotted in Fig. \ref{fig:coherence} (c) (green data points). These values now allow us to extract the transition lifetime $T_1$ from the fitting parameter $\eta$ shown in Fig. \ref{fig:g2} (d). The resulting values are given in Fig. \ref{fig:coherence} (c) as well. Averaging over the obtained $T_1$ values, we obtain the Fourier limit $2T_1$ (grey dashed line in Fig. \ref{fig:coherence} (c), with the error (grey shaded region) given by the standard deviation of $T_1$ values. For all but the highest excitation power, the coherence times are within the error bars of the Fourier limit, meaning that for resonantly generated single photons from this source, the Fourier limit is actually the lower bound of observed coherence times. Such Fourier limited emission has previously only been reported around 900 nm \cite{Kuhlmann.2015, XingDing.2016, Wang.2016, Somaschi.2016}.

Finally, we plot the values for $A$ as a function of Laser intensity, and compare them to the values obtained from the HOM measurements described above. As seen in Fig. \ref{fig:coherence}, the two methods give largely agreeing values. Theory predicts that the coherent fraction $F_{CS}$ depends on the driving intensity as follows \cite{Phillips.2020}:
\begin{equation}
F_{CS} = \frac{1}{1+\frac{\Omega^2}{\gamma}},
\end{equation}
where  $\gamma$ is the natural linewidth of the quantum dot. This model fits our data reasonably well [Fig. \ref{fig:coherence} (d)]. Residual deviations can be attributed to imperfect calibration of the HOM interferometer as well as drifts in the setup leading to differing effective excitation intensities, or also to phonon sideband contributions, which ultimately limit the coherently scattered fraction of photons \cite{Koong.2019, Brash.2019}. While a precise determination of this contribution is left for a future study, the high degree of elastic scattering as well as the $T_2$ times near the Fourier limit suggest that this process is of limited importance in our InAs/InP QDs.

\section{Indistinguishability of photons separated by 25 km of fibre}

Next, we make direct use of the minimal attenuation in fibre at the emission wavelength of our QDs and measure the indistinguishability by adding a 25-km fibre delay to one of the arms of the interferometer shown in Fig. \ref{fig:HOM} (a). At 0.173/km dB for the standard SMF-28 Ultra used, this attenuates the signal in the long arm by 4.37 dB or a factor 2.74 compared to the short arm. For comparison, the $\sim$ 4dB/km-attenuation in specialized fibre at 900 nm would still result in a signal attenuation by 10 orders of magnitude, making an indistinguishability measurement all but impossible. 

\begin{figure*}
 	\centering
 	\includegraphics[width=0.8\textwidth]{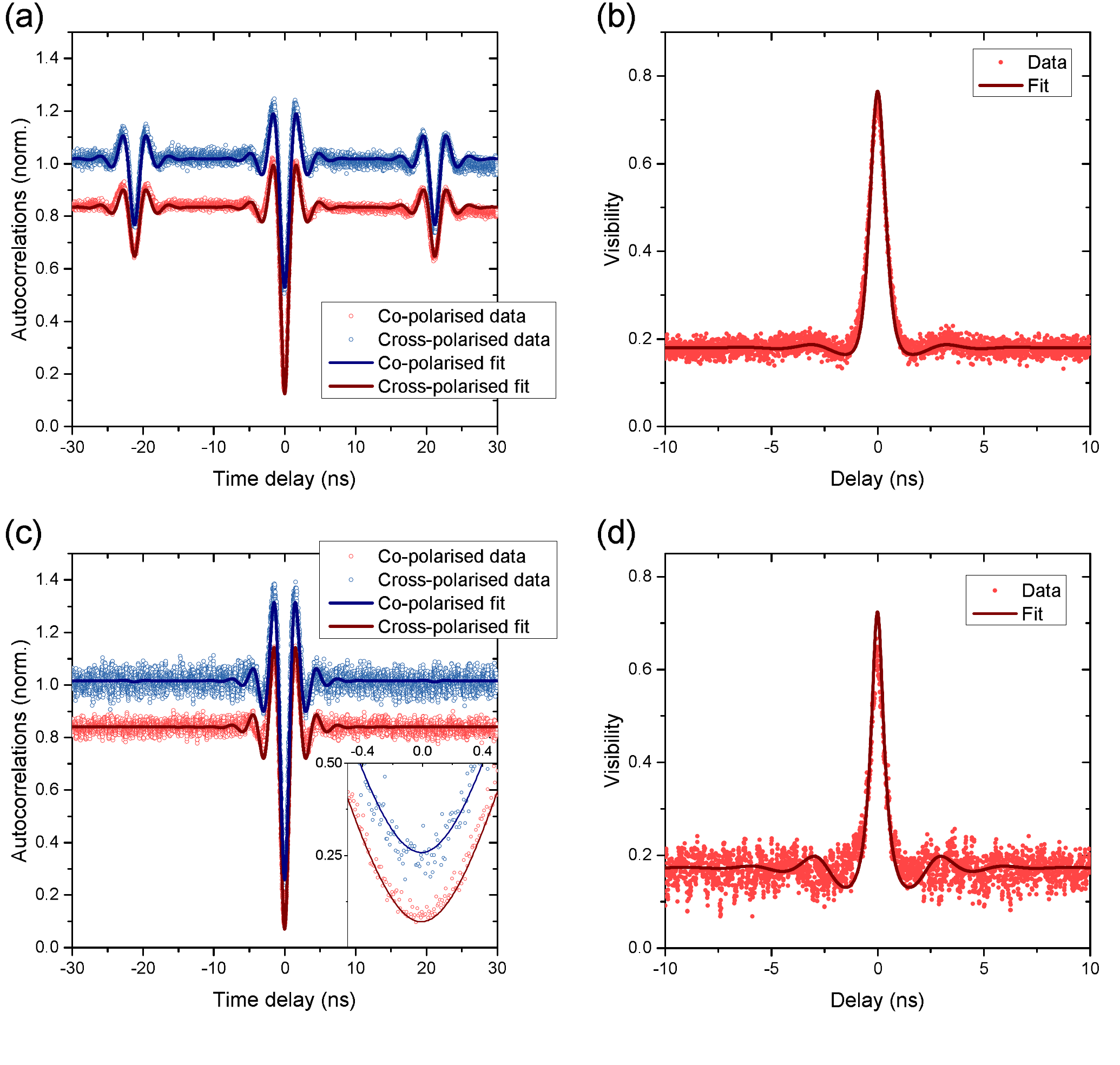}
 	\caption{\textbf{HOM measurements after 25 km of fibre} (a) HOM measurement for dot 2 without the long fibre delay. (b) Visibility calculated from (a). (c) Autocorrelation measurements passing one arm of the HOM interferomter through 25 km of fibre. (d) Visibility resulting from (c)}
 	\label{fig:25kmHOM}
 \end{figure*}

For the subsequent measurement, a separate QD was used. To establish a baseline, we first measure the indistinguishability without the extra delay, in the same configuration as above. Fig. \ref{fig:25kmHOM} (a) shows the measured correlations for the co- and cross-polarized cases, where the co-polarized data has been normalized taking into account the coherently scattered fraction of the light as above. The extracted visibility is shown in \ref{fig:25kmHOM} (b). At 0.77$\pm$0.06, the maximum is slightly higher than the values measured for QD 1, likely due to better experimental overlap of the mode functions. 

Next, we insert a 25-km fibre spool into the long arm of the interferometer. The measured relative attenuation of this arm is 5.8 dB and includes extra connectors as well as a switch for polarization calibration. The correlation measurements resulting from this configuration are shown in Fig. \ref{fig:25kmHOM} (c). It is immediately obvious that the side dips seen in the short-delay configuration have disappeared, as they have shifted out of our measurement window to a time delay of $\sim 95$ $\mu$s, corresponding to the extra distance traveled in the fibre. We can therefore no longer use the measured dip depth of the side dips to extract the coherent fraction, and have to rely instead on the short-delay measurement performed at a similar excitation power. We further note that the cross-polarized correlation curve at zero delay dips well below the 50\% mark expected for a balanced interferometer, seen in the inset to Fig. \ref{fig:25kmHOM} (c) showing a zoom around the zero-delay dips. This is due to the extra attenuation in one of the two arms, which makes correlations arising from two photons having both traveled the short arm dominant compared to the other contributions in Fig. \ref{fig:25kmHOM} (b). For the measured attenuation values, we expect a $g^{(2)}_{cross}(0)=0.33 \pm 0.03$, in reasonable agreement with the fitted value of 0.27$\pm$0.02. The resulting visibility is shown in Figure \ref{fig:25kmHOM} (d), and we obtain a value of 0.72$\pm$0.15. The extra fibre therefore results in a drop of visibility by about 6.5\%, meaning that even after $\sim$100 000 excitation cycles, the transition remains little affected by additional spectral wandering and other slow dephasing processes, demonstrating the stability of our QD transition on the timescale of $95$ $\upmu$s.

\section{Conclusion}


In conclusion, we have used HOM measurements as well as direct measurements of emission coherence to show that, except for very high driving powers, the coherence time of scattered photons is at least equal to the Fourier limit, and exceeds it considerably for low driving powers due to a large fraction of coherently scattered photons. For this particular QD, the Fourier limit for inelastically scattered photons is reached even without any additional Purcell enhancement or active feedback on the quantum dot. We further measure indistinguishability of photons emitted $\sim95$ $\upmu$s apart by guiding one of the photons through a 25-km fibre spool, a measurement infeasible with quantum emitters at lower wavelength. While spectral wandering processes affecting the QDs on the timescale of seconds are currently limiting the achieved visibilities, we expect that placing the QDs in doped structures similar to our earlier measurements \citep{Anderson.2021} will readily improve indistinguishability.

An outstanding challenge hampering network integration of C-band quantum dots is further the improvement of the limited extraction efficiency. Recent proposals integrating QD sources into circular Bragg gratings or micropillars structures in the telecom C-band show that coupling efficiencies into single mode fibres up to around 80\% are possible while at the same time also providing Purcell enhancement of factors 10-43 \cite{Barbiero.2022, Bremer.2022}. Seeing that under non-resonant excitation the investigated QD has a coherence time only about a factor three higher than the average in InAs/InP QDs \cite{Anderson.2021}, we expect that combining resonant excitation with appropriate photonic engineering will enable the majority of the QDs to perform at the Fourier level with high efficiency. Such a device will be a desirable hardware component for quantum network applications ranging from simple point-to-point quantum key distribution to distributed quantum computation tasks based on interference of entangled photons linking remote locations.

\begin{acknowledgments}
The authors gratefully acknowledge the usage of wafer material developed during earlier projects in partnership with Andrey Krysa and Jon Heffernan at the National Epitaxy Facility and at the University of Sheffield. They further acknowledge funding from the Ministry of Internal Affairs and Communications, Japan, via the project ‘Research and Development for Building a Global Quantum Cryptography Communication Network’.  L. W. gratefully acknowledges funding from the EPSRC and financial support from Toshiba Europe Limited. \\
%
%
\end{acknowledgments}

\appendix

\section{Setup}
\label{app:setup}

Our sample consists of a single layer of self-assembled, droplet epitaxy InAs quantum dots, grown in the centre of an InP cavity. The cavity is asymmetrical, with 20 bottom DBR pairs and 3 top distributed Bragg reflector pairs. This enhances the efficiency of the structure by directing emission away form the substrate. A 1 mm diameter, cubic zirconia solid immersion lens is attached to the top of the sample via 290 nm of HSQ. This also enhances the collection efficiency of the sample. 

\begin{figure*}[h]
 	\centering
 	\includegraphics[width=0.8\textwidth]{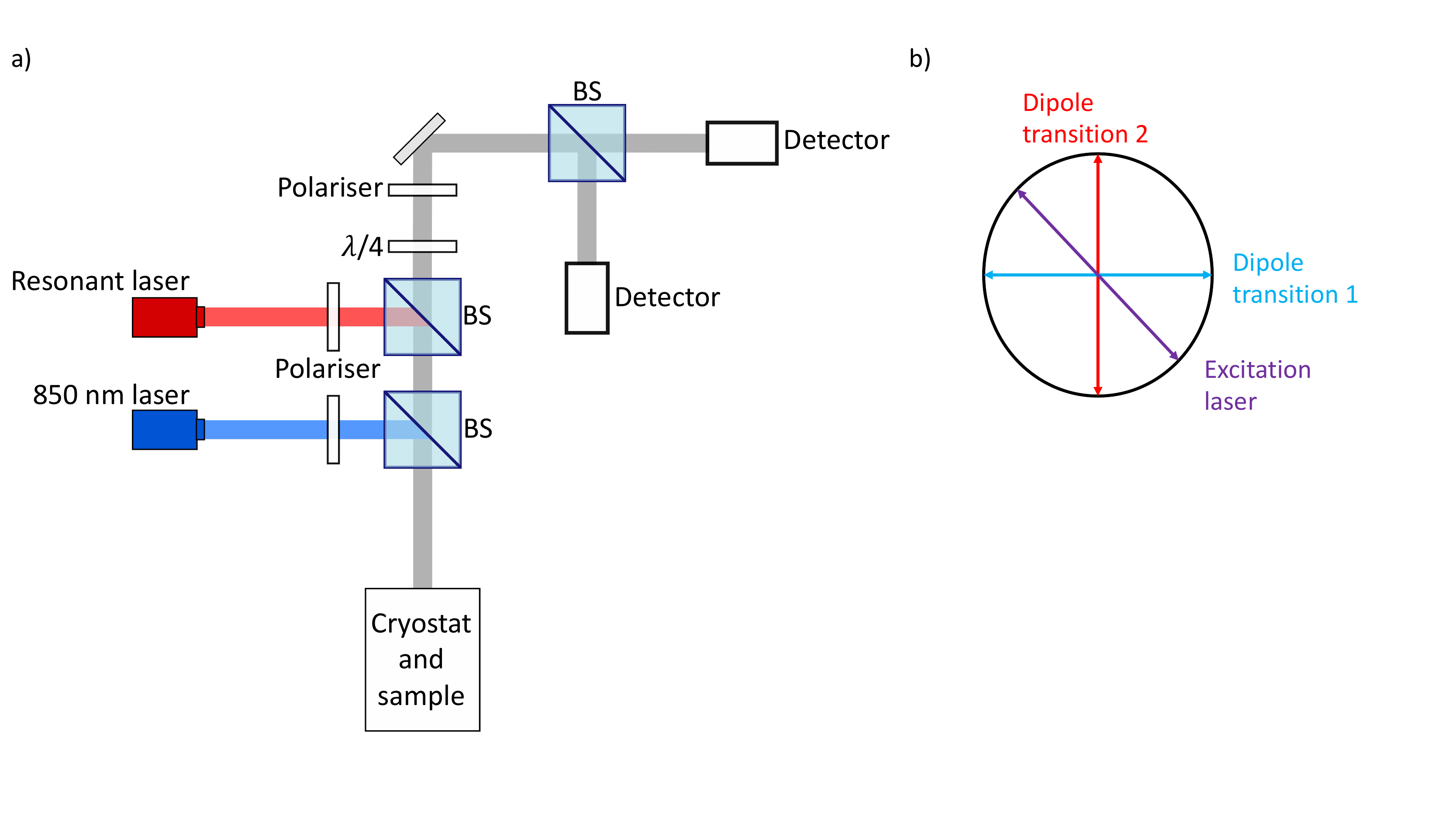}
 	\caption{\textbf{Schematic of the experimental setup} (a) Schematic of the main experimental setup. (b) Schematic showing the orientation of hte excitation laser relative to the two exciton dipole transitions.}
 	\label{fig:setupmain}
 \end{figure*}

Our main experimental setup is shown in Figure \ref{fig:setupmain}a. We used two lasers: a weak non-resonant laser (850 nm shown in blue) to measure photoluminescence from the quantum dot, while a narrow linewidth continuous wave laser scanned around 1532.2 nm resonantly excites the exciton transition. The resonant laser passes through a linear polarizer to ensure that the quantum dot is efficiently excited. This means that the input laser polarization is at approximately 45$^{\circ}$ with respect to the dipole transitions, which in turn are 90$^{\circ}$ apart. This achieves the best balance between high count rates and good laser coupling. The beam is then focused onto the cooled QD-SIL system within the cryostat via a microscope. A quarter waveplate and linear polarizer are placed in the detection path to allow for cross-polarization filtering. The collected photons then pass through either a HBT setup or a HOM setup, before being recorded by two SSPDs (detectors). The system is held at 10 K. A single exciton transition was used for all the data presented here. The photoluminescence spectrum of the QD under non-resonant excitation is shown in Figure \ref{fig:setupmain}b, with the exciton transition indicated. 

\section{HOM normalisation}
\label{app:HOMnorm}

\begin{figure*}[h]
 	\centering
 	\includegraphics[width=0.5\textwidth]{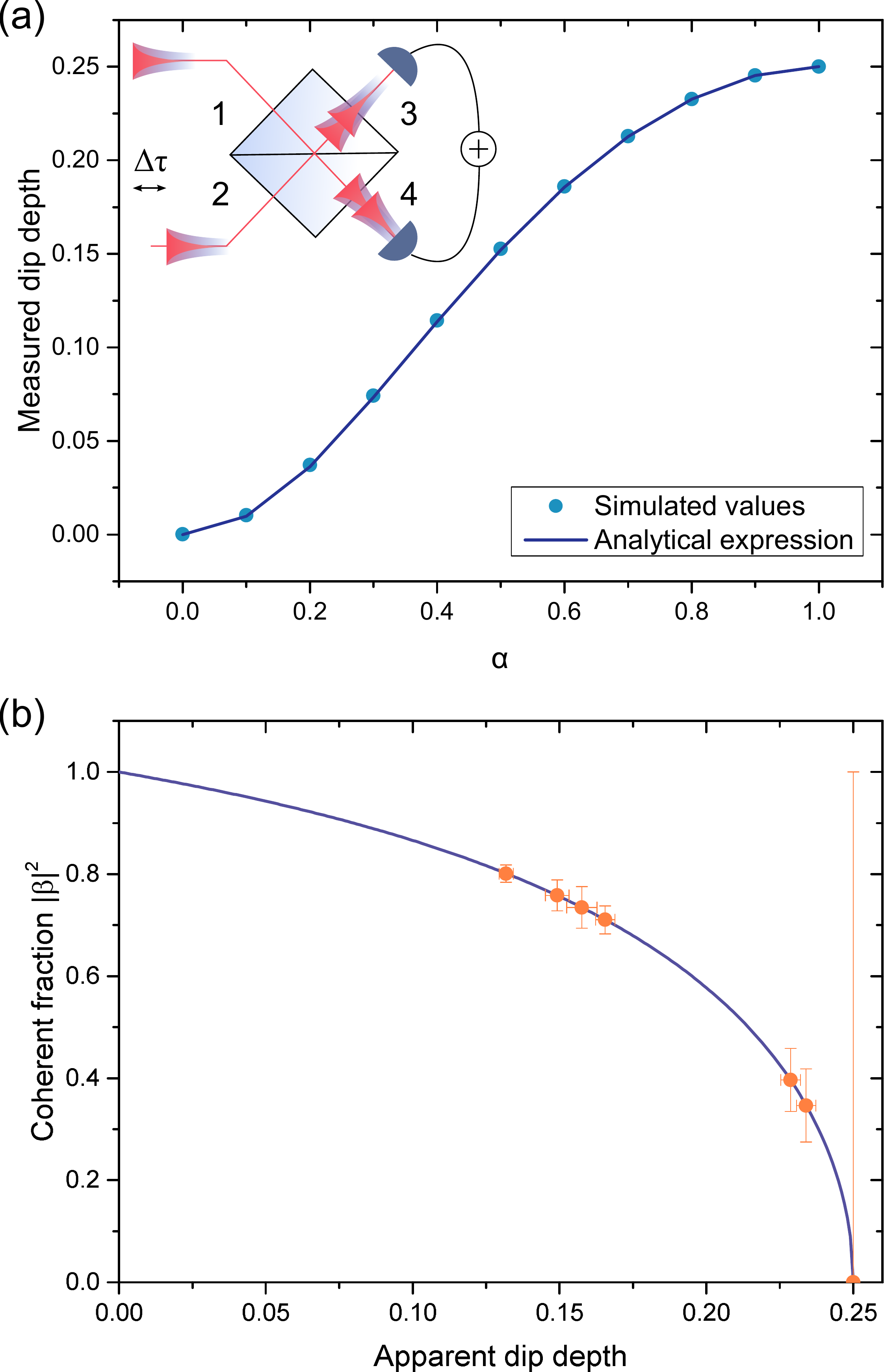}
 	\caption{(a) Measured dip depth calculated analytically from the model wavefunctions given in Eq. \ref{eq:HOMmodel}. The inset shows a schematic of the second beamsplitter in the HOM setup. (b) Theoretical expression for the coherent fraction $|\beta|^2$ as a function of dip depth (blue curve), and coherent fractions extracted from experimentally determined dip depths (orange data points).}
 	\label{fig:HOMmaths}
 \end{figure*}

To derive an analytical expression for the relationship between the incoherently scattered part of the mode function $\alpha$ and the observed dip depth, we explicitly calculate $P_{HOM}(\tau, \|)$ using the mode functions given in Eq. \ref{eq:HOMmodel} for inputs 1 and 2 at the second beamsplitter, as shown in the inset to Fig. \ref{fig:HOMmaths}. Following Refs \cite{Legero.2003, Patel.2010, Anderson.2021}, the probability of a joint click of the two detectors for indistinguishable light is given by
\begin{equation}
P^{joint}_{\|}(t, \tau, \Delta \tau) = \frac{1}{4}\left|\xi_1(t+\tau-\Delta\tau)\xi_2(t)-\xi_1(t-\Delta\tau)\xi_2(t+\tau)\right|^2,
\end{equation}
where $\tau$ gives the time delay between the clicks and $\Delta \tau$ accounts for the different arrival times of the photons at the beamsplitter. Integrating over the absolute time $t$ and $\Delta \tau$ gives
\begin{eqnarray}
P_{HOM}(\tau, \|) &=& \int_{-\infty}^{+\infty} \int_{-\infty}^{+\infty}  P^{joint}_{\|}(t, \tau, \Delta \tau) \,\mathrm{d}t\,\mathrm{d}\Delta\tau\\
&=&\frac{1}{2}\left(1+|\alpha|^4 e^{-\frac{2}{T_1}|\tau|}+2 |\alpha|^2 |\beta|^2 e^{-\left(\frac{1}{T_1}+\frac{1}{T_2})\right)|\tau|}+|\beta|^4 e^{-\frac{2}{T_2}|\tau|}\right).
\label{eq:totaldip}
\end{eqnarray}

Entering this expression into Eq. \ref{eq:corrs}, we can calculate the the expected correlations, which are shown in Fig. \ref{fig:HOMmodel}. From these curves, we numerically extract the measured dip depth as a function of $\alpha$, shown as blue circles in Fig. \ref{app:HOMnorm}. We also calculate the measured dip depth $D$ analytically, from the parts of Eq. \ref{eq:totaldip} which contribute to the measured dip, $C = |\alpha|^4+2|\alpha|^2|\beta|^2$, and the parts which do not, $NC = |\beta|^4$. We obtain the analytical expression
\begin{equation}
D = \frac{1}{2}\frac{2|\alpha|^2-|\alpha|^4}{2|\alpha|^2-|\alpha|^4+1},
\end{equation}
shown as the blue curve in Fig. \ref{fig:HOMmaths} (a). This expression results in a quadratic equation to determine $|\alpha|^2$ from a measured $D$:
\begin{equation}
|\alpha|^2 = 1 -\frac{\sqrt{8D^2-6D+1}}{1-2D}.
\end{equation}
We obtain $D$ from the co- and cross-polarized correlation measurements after deconvolution with the detector response and fitting with Eq. \ref{eq:corrs}. To account for laser background and imbalanced intensities in the interferometer, the co-polarized dip depth is normalized to the cross-polarized dip depth. The measured $D$ then determines the coherent fraction $|\beta|^2 = 1-|\alpha|^2$. The theoretical expression for $|\beta|^2$ as well as the experimentally determined dip depths and extracted coherent fractions are shown in Fig. \ref{fig:HOMmaths} (b).\\
To re-normalize the data, we scale the measured dip depth to the theoretically expected dip depth based on the calculated coherent and incoherent fractions. Given that $NC+C=1$ by construction, the expected dip depth for a given coherent fraction $|\beta|^2$ is 
$(1-|\beta|^2)\times D_{cross}$, where $D_{cross}$ is the cross-polarized dip depth. $D_{cross}=0.25$ in the ideal case of no background emission and a balanced interferometer, and in our experiments without the long delay is found to be $D_{cross}=0.237\pm0.003$. We therefore apply the following normalization factor to the co-polarized data:
\begin{equation}
N = \frac{(1-|\beta|^2)D_{cross}}{D}.
\end{equation}
This factor is in addition to the Poissonian normalization applied to the cross-polarized data.


%
\bibliographystyle{naturemag} 
%

\end{document}